\documentclass[preprint,proceedings]{rmaa}
\suppressfulladdresses % by popular request
\usepackage{paralist}
\usepackage{psfrag,color}

\SetYear{2002}
\SetConfTitle{ Circumstellar Media and Late Stages of Massive Stellar Evolution

}
\title{Massive binaries, Wolf-Rayet stars and supernova progenitors.} 

\author{
  J. J. Eldridge\altaffilmark{1} }

\altaffiltext{1}{Astrophysics Research Centre, Department of Physics \& Astronomy, Queen's University, Belfast, BT7 1NN, United Kingdom. (j.eldridge@qub.ac.uk)}

\shortauthor{Eldridge}
\shorttitle{Massive binaries, WR stars and SN progenitors.}

\listofauthors{J.J. Eldridge}
\indexauthor{J.J. Eldridge}

\abstract{Binary stars are important for a full understanding of stellar evolution. We present a summary of how predictions of the relative supernova rates varies between single and binary stars. We also show how the parameter space of different supernova types differs between single and binary stars. We then consider an important question of how to infer a supernova progenitor's properties from pre-explosion imaging and present rescent work of producing synthetic colours for our stellar models to make a direct comparison with any detections or limits obtained on supernova progentiors from pre-explosion images.}

\addkeyword{Stars: binaries}
\addkeyword{Stars: circumstellar material}
\addkeyword{Stars: Pre-main sequence}
\addkeyword{Stars: Mass loss}

\begin{document}
% Typeset article header
\maketitle

\section{Introduction.}

Binary stars are a vital part of stellar evolution if we wish to gain a full understanding of all the possible pathways a star may take from its birth to its death. However binaries are normally ignored due to the difficulty in covering the large possible initial parameter space that has to be studied. For single stars we only need consider initial mass, initial metallicity and mass-loss rates (also rotation rate if stellar rotation is considered); binary stars add secondary mass and initial orbital separation (or its equivalent the orbital period). Both of these extra variables have a large range, especially the initial separation that could be only a few stellar radii up to many thousands of stellar radii. To fully understand binary evolution this entire range has investigated. Therefore where a set of single models contain a few hundred models a complete binary set contains a few thousand at least.

We discuss the implications of binaries for supernovae (SNe) and their progenitors. First we discuss some general details of our binary models and then we discuss how mixing populations of single and binary stars can obtain agreement with observed supernova rates. We then move onto considering supernova progenitors, especially how to predict their appearance in pre-explosion images and therefore gain more information on the nature of observed SN progenitors. First we consider red supergiants (RSGs) as the progenitors of type IIP supernovae and then Wolf-Rayet (WR) stars; these are massive star that have lost their hydrogen envelopes to become naked helium-stars and are the progenitors of type Ibc SNe. We show how with predicted UBVRI colours we are able to place very tight constraints on these types of progenitors from current data.

\section{Stellar Models}

\begin{figure}[!t]
  \includegraphics[width=\columnwidth]{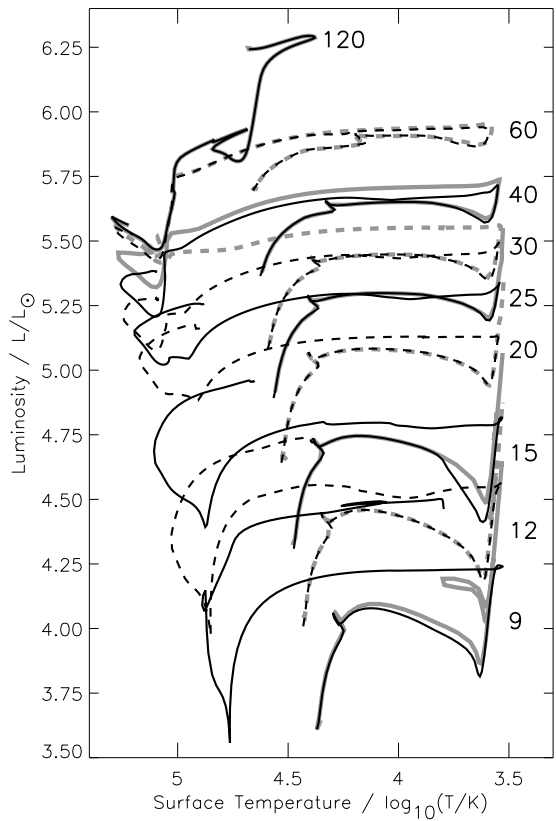}
  \caption{HR diagram for single stars (grey tracks) and binary stars (black tracks). The labels on the right hand side indicate the initial mass of the stars in M$_{\odot}$.}
  \label{hr}
\end{figure}

\begin{figure}[!t]
  \includegraphics[width=\columnwidth]{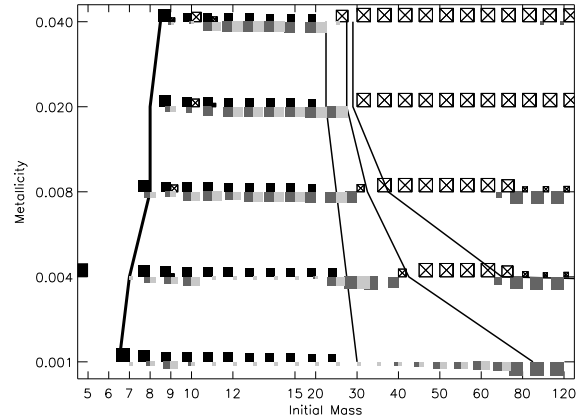}
  \caption{Parameter space over which the progenitors for different SN types occur. Type IIP SNe occur where there is a black box, the light grey boxes indicate the region of type IIL SNe, the dark grey boxes are where Ib SNe occur and the boxes with crosses are the region of type Ic SNe. The area of each box is roughly proportional to the relative fraction of SNe for each initial mass and initial metallicity. The solid black lines indicate the single star SN progenitors parameter space. The regions indicate from left to right are no-SNe, IIP SNe, IIL SNe, Ib SNe and Ic SNe.}
  \label{heger}
\end{figure}

Our stellar models are constructed using the Cambridge STARS code (Eldridge \& Tout 2004 and references therein). The single star models are calculated using our preferred mass-loss rates (scheme D in Eldridge \& Vink 2006). We can see in figure \ref{hr} the evolution tracks of single and binary stars. This figure shows a wide range of possible progenitors. The binary evolution scheme is a simple treatment based on \citet{HPT02} and is essentially that described in Eldridge (2004). The binary code does not follow rotation.

Single stars up to about 25M$_{\odot}$ end their lives as red supergiants while stars above this limit become Wolf-Rayet stars by losing their hydrogen envelopes when they become RSGs or luminous blue variables (LBVs) for the most massive stars. The binary star tracks however are very different. In figure \ref{hr} we have selected binary models that interact and therefore they lose their hydrogen envelopes to become helium stars. This leads to an interesting question, when is a helium star a Wolf-Rayet star? It appears that stars initial more massive than about 15-20M$_{\odot}$ are WR stars while the lower mass stars seem to be quite different from WR stars and have larger radii, cooler envelopes and tend to retain more helium than the more massive stars. They are helium giants and will be a common source of type Ib SNe.

It is important to note that there is a clear division been that stars greater than about 25M$_{\odot}$ that have the most severe winds and become WR stars even as single stars and therefore mainly produce type Ic SNe. Stars less massive than this need something extra to remove their hydrogen envelopes and the remaining variety for SN types (IIP, IIL, IIb, IIn and Ib) come from this mass range of progenitors. There are undoubtedly exceptions to this simplification and there are some bizarre binary progenitor scenarios that can be created but on average this should be our conclusion.

In figure \ref{heger} we plot were the different SN types occur over initial mass and initial metallicity space for binary stars. The mass ranges change with initial metallicity slightly for binary evolution. This is maybe unexpected but is due to binary interactions only being able to remove a certain amount of the hydrogen envelope with the remaining amount removed by the stellar winds. Therefore at low metallicity with the weaker winds it becomes more difficult to remove every scrap of hydrogen from the star thus the parameter space where type Ibc SN occur does reduce.

One final interesting result is that the Ic SNe are concentrated in the region where single stars will produce Ic SNe. Some of the lower mass stars do produce some Ic SNe but the majority of the SNe in this mass range will be other core-collapse SN types.

\section{Supernova Rates}
With predictions of which SN different binary stars might produce we can make predictions of the relative rates of SNe and compare these to observed rates. This can reveal whether the mass-loss our models predict over the lifetimes of the stars is correct or not. Essentially if mass-loss is more severe there will be more type Ibc SNe than those of type II. Therefore in general we should expect fewer type Ibc SNe at low metallicities because winds are expected to become weaker at lower metallicities (Vink et al. 2001 and reference therein).

\begin{figure}[!t]
  \includegraphics[width=\columnwidth]{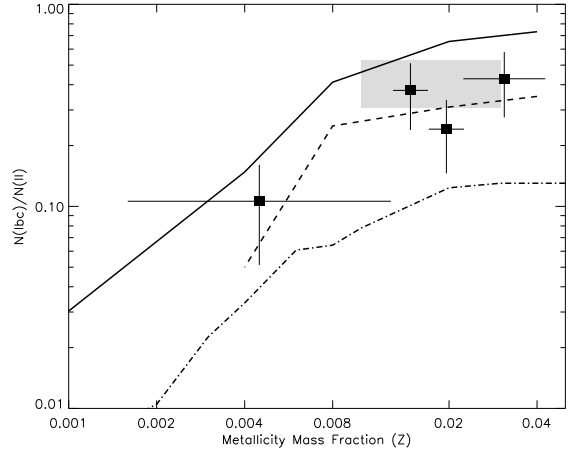}
  \caption{Rates versus metallicity. The observed points are from \citet{snevsZ}, the grey box is from \citet{smartt2007}. The solid line is from our binary models, the dashed line is predicted ratios from \citet{mm2004} and the dash-dotted line is from out single star models.}
  \label{rates}
\end{figure}

In figure \ref{rates} we see this is indeed the case from the observations of \citet{snevsZ}. We also plot on this the relative rates from \citet{smartt2007} calculated from a local volume limited survey. These rates are in good agreement with the earlier rates around solar metallicity.

Comparing predicted rates we see that at first glance the single star rotating models of \citet{mm2004} agree well with the observed rates while our single and binary star predicted rates are too low and too high respectively, but in reality we would have a mixture of single and binary stars. Mixing these systems together we would find agreement with the observed rates. Importantly in the binary case the relative fractions of type Ibc SNe does decrease with metallicity.

We have also compared our models to various stellar population ratios such as the blue supergiant (BSGs) to RSGs ratio, RSGs to WR star ratio and WR to O star ratio and find that in general a mix of single stars and binary stars do lead to an agreement with observations. However agreement is not exact because our binary model is simple and does not yet include tides or rotation. There is also another problem of mapping observed stellar types onto stellar models. For example because a stellar model has the parameters that would make it counted as a WR star would it be observed as such? We find our predicted ratios at low metallicities can be strongly affected by the lower luminosity cut-off where we take a star to be a helium star rather than a WR star.

\section{What do SN progenitors look like?}

Stellar evolution codes produce very simple outputs that describe the appearance of a star for comparison to observations. These are in essence the surface temperature and luminosity. These two values are only truly obtainable via spectrometry however this is not always available for SN progenitors apart from the exceptional case of the progenitor of SN 1987A. It is more common that photometry exists but it is difficult to directly connect this to a stellar models. To infer stellar parameters from a photometry detection a bolometric correction must be assumed. This means the stellar type must be assumed and a systematic error is introduced.

Here we suggest that a less error prone method is to calculate the observable photometric properties of our stellar models. Our method is based on that of \citet{LS01}. We use the atmosphere library of \citet{basel3} with the photometric filter functions of \citet{ubvribands} and \citet{jhkbands} to calculate the $UBVRIJHK$ magnitudes. Then by interpolating of $\log g$ and $T_{\rm eff}$ we produce the colours and magnitudes in these bands in various useful combinations.

Importantly for the first time we have extended the calculation of the colours to the Wolf-Rayet phase of evolution by using the atmosphere models of the Potsdam group \citep{wrspec1,wrspec2}. With the broad-band magnitudes known for the entire evolution for our single and binary star models we know the broad range of colours possible for supernova progenitors and now we can analyse the progenitors in greater detail and move their study beyond simply identifying the progenitors and perhaps limiting the various evolution paths of massive stars.

\section{RSGs}

RSGs are the progenitors of type IIP SNe as they require massive and extended hydrogen envelopes to produce the plateau phase in the lightcurve this occurs because the photosphere moves inwards in ejecta mass as the ejecta expands in radius. RSGs are very luminous and very cool with luminosities greater than around $10^{4.3}L_{\odot}$ and surface temperature cooler than around 5000K. Therefore their magnitudes are greatest in the RIJHK bands. This is because the peak of the black-body spectra is similar to the wavelength these bands cover. There is an important advantage of the RSG colours being weighted towards the red and near infra-red that any mass determinations are less sensitive to dust \citep{dust}.

There is an interesting result of estimating the final colours of our models for stars near the minimum mass for a SN to occur. This is because these stars may go through second dredge-up. Dredge-up is a process in stellar evolution that occurs between the main burning stages. First dredge-up occurs after core hydrogen burning as the core contracts before helium burning ignites. The hydrogen envelope becomes convective and this convective zone penetrates deep into the core dredging-up material that has been processes by the central nuclear reactions. Therefore the abundance of helium and nitrogen are boosted while hydrogen, carbon and oxygen abundances decreased.

Second dredge-up occurs after core helium burning and the convective envelope penetrates much more deeply; pushing the hydrogen burning shell into close proximity with the helium burning shell. This arrangement is unstable and leads to burning occurring in pulses. The hydrogen shell burns out until there is enough helium for the helium burning to occur and all the helium is rapidly burnt. Afterwards the hydrogen shell again burns outwards and the process repeats.

The more massive class of these stars ($\ga 6 M_{\odot}$) experience carbon burning during or after second dredge-up. The leads to the formation of an oxygen-neon-magnesium (ONeMg) cores within the helium and hydrogen burning shells. We note that the carbon burning shell extinguishes after the ONe core is formed and never reignites.

These massive-AGB stars may give rise to a SN. The core can continue to grow in mass until it reaches the Chandrasekhar mass where the electron degeneracy pressure in the core can no longer provide support and it therefore collapses to a neutron star, if oxygen burning ignites it cannot halt collapse unlike carbon burning in a CO white dwarf the reaches the Chandrasekhar mass. Observationally the SNe may be very low energy but otherwise identical to other type IIP SNe. See \citet{ETsne}, \citet{tagb} and Langer et al. this volume for further details.

\begin{figure}[!t]
  \includegraphics[angle=270,width=\columnwidth]{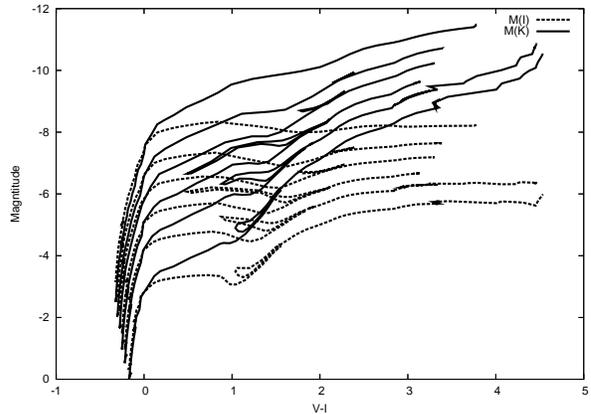}
  \caption{The I and K band magnitude versus the V-I colours for stars with initial masses of 5, 7, 9, 12, 15 and 20 M$_{\odot}$ from top to bottom. The solid lines are the I band magnitude and the dashed lines are K band magnitudes. The 5 and 7 M$_{\odot}$ tracks end their lives as AGB stars and have larger I-K colours than the RSGs.}
  \label{magb}
\end{figure}

Whether these progenitors exist or whether some other type of evolution occurs to prevent them producing supernova is of great interest. The most secure method to observe these stars will be to observe the progenitor. AGB stars are normally cooler than RSGs but about 1000K this leads to their higher magnitudes in the JHK bands. For example in figure \ref{magb} we see that during the evolution of stars that while the I band magnitude is fairly constant for a specific initial mass the near infrared K band varies much more for the lower mass stars that become AGB stars. Therefore for any AGB star SN progenitor the I-K colour will be very different to that for a normal RSG.

The last important detail we should remember though is what ever observations exist for progenitors of type IIP perhaps the best band to use is the I band as it is not effected by variation in the surface temperature and mainly varies with the initial mass of the tracks.

\section{WR}

Currently we can consider the problem of type IIP as not completely solved but better resolved. However mystery still enshrouds the remaining SN types or IIL, IIb, IIn and Ib/Ic. These final two SN types are of great interest as conclusions we can draw on the progenitors of these objects will have implications for Gamma-ray bursts (GRBs) as some long GRBs have been observed to be associated with type Ic SNe \citep{sngrb01,sngrb02}.

There is another problem that it is important and that is whether the progenitors of type Ib's and type Ic's are the same. Type Ic's have no hydrogen or helium while type Ib's still have helium in their envelopes. Stellar models indicate that both types have two types of progenitors, low mass and high mass.

For type Ib progenitors there is a narrow range for massive stars around 25-27M$_{\odot}$ which are WR stars that retain a large amount of helium. The more massive stars have stronger winds and these can remove enough helium from the star so that it would not be observed in the supernova and therefore produce a type Ic progenitor.

The stars less massive than about 25M$_{\odot}$ however can only produce a type Ibc supernova if they are in an interacting binary and their hydrogen envelopes are removed. The winds in these stars are weaker so once the hydrogen envelopes are removed then the stars will retain helium and produce a type Ib SNe. However it is possible that the interactions continues, for example helium stars less massive than about 5M$_{\odot}$ become helium giants with extended envelopes. These can interact for a second time producing very low mass type Ic progenitors. The initial parameter space is shown in figure \ref{heger}.

\begin{figure}[!t]
  \includegraphics[width=\columnwidth]{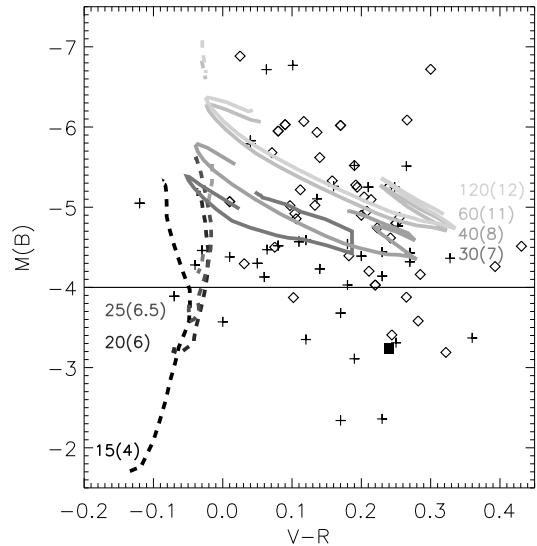}
  \caption{WR tracks of B band magnitude versus V-R colour. The dashed tracks are for the WN phase of evolution while the solid lines are for WC evolution. The labels indicate the initial mass and the final mass in M$_{\odot}$. The observed points are taken from Massey's catalogue of M31, the diamonds are WN stars, the crosses WC stars and the black square a WO star. The thin black line is the upper limit for the progenitor of 2002ap from Crockett et al., in prep.}
  \label{wrcols}
\end{figure}

There are a number of type Ibc SNe where limits exist on the progenitors, however most of these have luminosity limits about the same level as the most luminous WR stars known (see Figure \ref{wrcols}). However there is one SN, 2002ap, where the luminosity limits are much lower and we can exclude most of the range of observed WR stars from the range of possible progenitors as we show in figure \ref{wrcols}.

In the same figure we also plot the tracks of the WN and WC phases of evolution. From here we see that all the WC models are above the luminosity limit and the WN tracks are below the luminosity limit. The mass of the progenitor has been inferred from modelling the SN lightcurve and spectra at around 5-6M$_{\odot}$ \citep{02ap}. Therefore while the SN was a type Ic and should be a WC rather than a WN progenitor, WR stars with a mass the same as that estimate from the SN observations are below the luminosity measurement and are therefore consistent. It may also indicate that we may require stronger mass-loss during some phases of evolution to produce lower mass WC stars.

\section{Conclusions \& future work}
Binaries are important to include in our studies of massive stars and SNe. They increase the range of possible SN progenitors and evolutionary pathways. We also see that the study of supernova progenitors is now evolving past only identifying them but trying to infer much more information on the evolutionary structure of the star before any SN. Another important reason to study binary evolution is that in the future there will be more SN discovered as new searches begin. This will lead to us discovering more unusual SN and binary evolution may be required to explain many of these.

\end{document}